\documentclass[reprint,amsmath,amssymb,aps,prl]{revtex4-2}
\usepackage{graphicx}
\usepackage{dcolumn}
\usepackage{bm}
\usepackage{blindtext}
\usepackage{xcolor}
\UseRawInputEncoding

\newif\ifarxiv
\arxivtrue 

\begin{document}

\preprint{APS/123-QED}

\title{Origin of correlated isolated flat bands in copper-substituted lead phosphate apatite}

\author{Sin\'{e}ad M. Griffin$^{1,2}$}
\affiliation{$^{1}$Materials Sciences Division, Lawrence Berkeley National Laboratory, Berkeley, California, 94720, USA}
\affiliation{$^{2}$Molecular Foundry Division, Lawrence Berkeley National Laboratory, Berkeley, California, 94720, USA}

\date{\today}

\begin{abstract}

A recent report of room temperature superconductivity at ambient pressure in Cu-substituted apatite (`LK99') has invigorated interest in the understanding of what materials and mechanisms can allow for high-temperature superconductivity. Here I perform density functional theory calculations on Cu-substituted lead phosphate apatite, identifying correlated isolated flat bands at the Fermi level, a common signature of high transition temperatures in already-established families of superconductors. I elucidate the origins of these isolated bands as arising from a structural distortion induced by the Cu ions and a chiral charge density wave from the Pb lone pairs. These results suggest that a minimal two-band model can encompass much of the low-energy physics in this system. Finally, I discuss the implications of my results on possible superconductivity in Cu-doped apatite.
\end{abstract}

\maketitle

\section{Introduction}

High-T$_C$ superconductors are arguably the holy grail
of condensed matter physics with huge potential applications for an energy-efficient future. The first class of superconductors that were considered to be high-T$_C$ were the cuprates which were discovered by Bednorz and M\"{u}ller in 1987~\cite{Bednorz_et_al:1986}. The cuprates have been subsequently followed by several new classes including the Fe-pnictides in 2008~\cite{Kamihara_et_al:2008} and the nickelates~\cite{Li_et_al:2019}.

While significant strides have been made in the discovery and understanding of high-T$_C$  superconductors, and we continue to unearth novel examples within established classes~\cite{Pan_et_al:2022}, a definitive roadmap to achieving room-temperature T$_C$ under ambient pressures has remained elusive. Common to many of these high-T$_C$ superconducting families are strongly correlated bands which can give rise to unconventional mechanisms for Cooper pair formation~\cite{Khodel_et_al:1990,Flatbands}, and proximity to multiple competing interactions such as antiferromagnetism, charge
density waves and spin-density waves. These phases can compete or coexist with superconductivity where fluctuations between these states are believed to play a significant role for achieving high-T$_C$. Searching  for these features in new materials systems is therefore a promising route for finding new classes of high-T$_C$ superconductors. For instance, the nickelate superconductors were originally predicted in theory by Anisimov, Bukhvalov and Rice as an analogy to the cuprate superconductors~\cite{Anisimov_et_al:1999}. Similar approaches have also been proposed to selectively design a material with the sought-after isolated d-manifold that is associated with strong correlations~\cite{Griffin_et_al:2016}, and have inspired high-throughput searches for good candidate materials, further expanding the horizons of high-Tc superconductivity~\cite{Isaacs/Wolverton:2019}.

The recent report of possible room temperature superconductivity at ambient pressures in Cu-substituted apatite (also known as `LK99')\cite{RoomT,RoomT2} motivates the need for a thorough understanding of the structure-property relationships in these compounds to begin to unravel their potential correlated physics. 
In this Letter, I use \textit{ab initio} calculations to elucidate
the key competing interactions in Cu-doped apatite at
the mean-field density functional level.

\section{Methods}
I used the Vienna Ab initio Simulation Package (VASP) \cite{Kresse1993,Kresse1994,Kresse1996,Kresse1996a} for all density functional theory (DFT) calculations with full calculation details given in the SI. I applied a Hubbard-U correction to account for the underlocalization of the Cu-$d$ states. I tested values of U between 2 eV and 6 eV, finding my results were similar for all values calculated. The results in the main text are for U = 4 eV which gives lattice parameters within 1\% of experiment~\cite{RoomT,Bruckner_et_al:1995}. 

\section{Results}
\subsection{Structural Properties}
\begin{figure}
\includegraphics[width=0.49\textwidth]{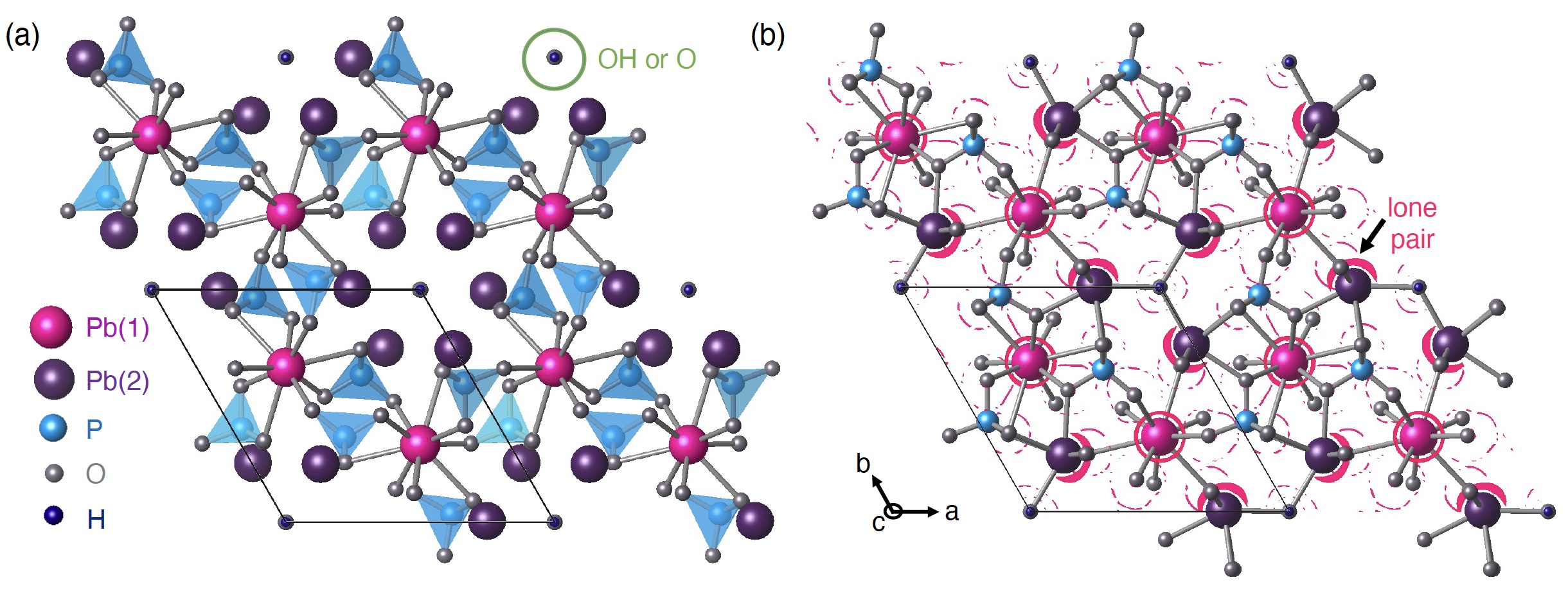}
\caption{(a) Lead-phosphate apatite structure with two inequivalent Pb sites as described in the main text. Columns of O or OH sit in the center column defined by Pb(2) hexagonal structure.  (b) Calculated electronic localization function for Pb$_{10}$(PO$_{4}$)$_{6}$OH$_{2}$. Oxygens surrounding Pb(2) are repelled by the lone pair.} 
\label{fig1}
\end{figure}

Apatites are materials with the general formula A$_{10}$(TO$_{4}$)$_{6}$X$_{2\pm x}$, where A = alkaline or rare earth metal; T = Ge, Si, or P; and X = halide, O, or OH. The name `apatite' derives from the Greek \textit{apat\={e}} meaning `deceit' as a result of the diverse range of forms it can take~\cite{Roycroft/Cuypers:2015}. Here I consider the lead-phosphate apatite Pb$_{10}$(PO$_{4}$)$_{6}$(OH)$_{2}$. Taking its structure reported from X-ray diffraction in Ref. \cite{Bruckner_et_al:1995} as the starting point, its structure following a full optimization is depicted in Fig.~\ref{fig1}. It adopts the typical crystal structure of various apatite chemistries, namely it forms a network comprising PbO$_{6}$ prisms that are corner shared with PO$_{4}$ tetrahedra. I refer to these Pb as Pb(1), in keeping with the convention in literature~\cite{Peet_et_al:2017}. This framework is filled with Pb$_{6}$(OH)$_{2}$ where the (OH)$_{2}$ forms a chain in the center of a hexagonal structure defined by these Pb(2) atoms. While here I consider OH$^{-1}$ filling the hexagonal column, I obtained similar results for O only in the column. Typically apatites adopt the hexagonal $P6_{3}/m$ space group -- here our resulting structure has $P6_{3}$ owing to the breaking of reflection from the OH molecule ordering. However, small structural deviations from the $P6_{3}/m$ commonly occur depending on its constituents.

While the Pb(1) forms the overall framework with the PO$_{4}$ tetrahedra, the Pb(2) play a crucial role in Pb-O connectivity and polyhedra tilts throughout the structure. In fact, while both Pb(1) and Pb(2) possess 6s$^{2}$ lone pairs, only the latter is stereochemically active. I verify this by calculating the electronic localization function (ELF) as shown in Fig.~\ref{fig1}(b), finding that the Pb(2) lone pairs form a chiral arrangement that make an angle of $\approx105^{\circ}$ with the \textit{a}-axis (see SI for more details). The arrangement of the lone pairs sets the resulting oxygen coordination of the Pb(2), namely its six oxygens are arranged asymmetrically as a result of their repulsion from the lone pair forming a chiral charge density wave. Since these oxygen are corner shared with the PO$_{4}$, the structural distortions associated with the Pb(2) lone pair activity propagate throughout the structure. Such an effect has been discussed previously in various apatite materials, noting that the lone pair ordering differs depending on the specific system \cite{Peet_et_al:2017, Antao_et_al:2018}.

\begin{figure}
\includegraphics[width=0.49\textwidth]{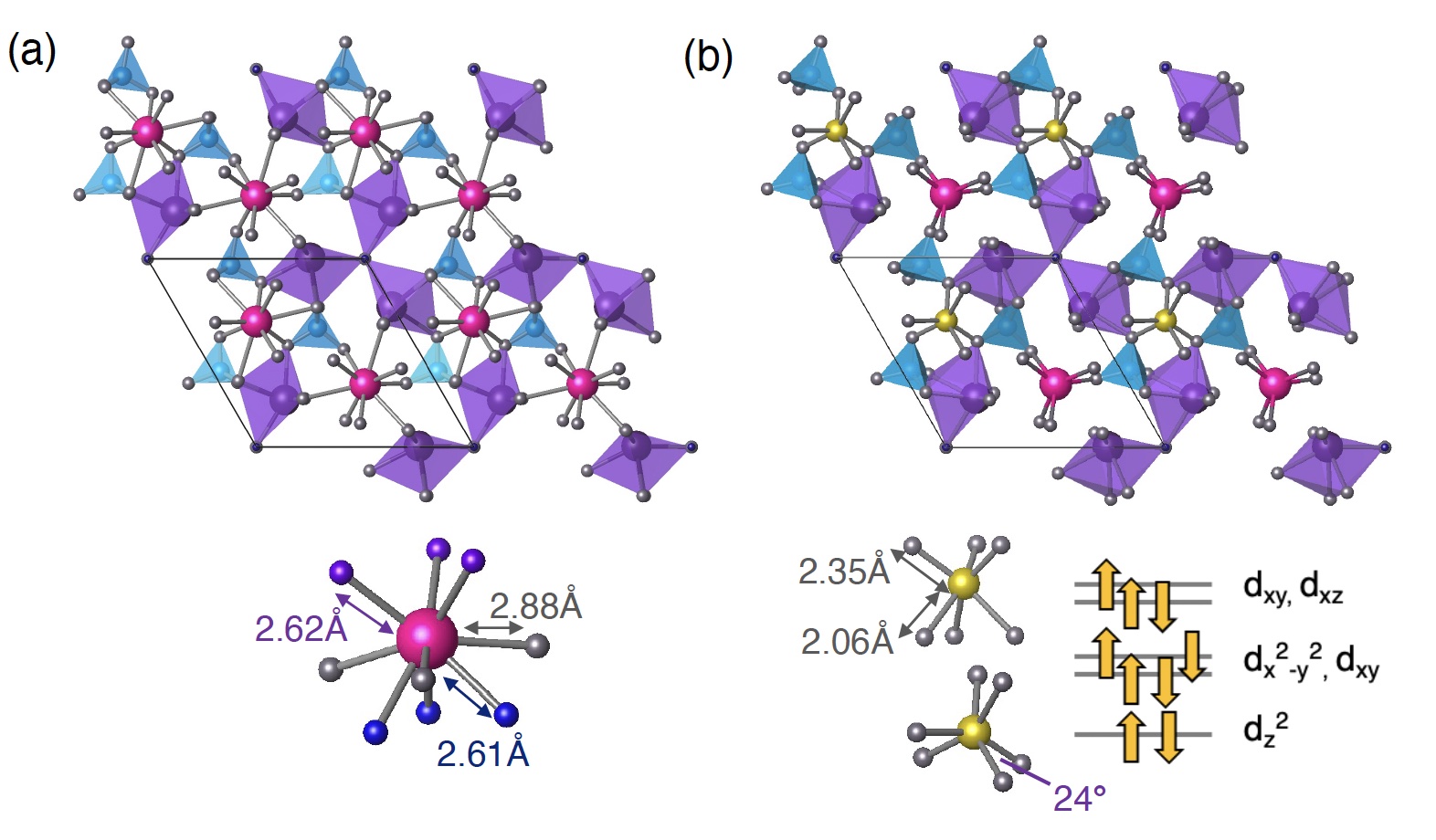}
\caption{(a) Lead-phosphate apatite structure showing nine-coordinated Pb(1) sites. (b) Cu-substituted structure showing six-coordinated Cu and Pb(1) sites with distorted trigonal prism coordination with two different bondlengths are a rigid twist of $\approx24^{\circ}$ between the upper and lower triangles. A cartoon of the resulting crystal-field diagram for Cu-$d^9$ is given on the right.} 
\label{fig2}
\end{figure}

I next consider the substitution of Cu on a Pb(1) site resulting in CuPb$_{9}$(PO$_{4}$)$_{6}$(OH)$_{2}$, with the fully optimized structure shown in Fig.~\ref{fig2}(b). I find several changes to the structure with the inclusion of Cu. Firstly, all lattice parameters decrease with $a$ going from 9.875~\AA\ to 9.738~\AA\, and $c$ going from 7.386~\AA\ to 7.307~\AA. While my calculated lattice parameters agree well with those reported in Ref.\cite{RoomT}, I find a greater structure collapse than their report of $a$ going from 9.865~\AA\ without Cu to 9.843~\AA\ with Cu, and $c$ going from 7.431~\AA\ without Cu to 7.428~\AA\ with Cu. Interestingly, I find Cu substitution results in a global structural distortion that results in a change in coordination from nine to six (with a cutoff of $3$ \r{A}) not only for the Cu on the Pb(1) site, but also for all other Pb(1) sites. This results in a modification of the polyhedral tilts throughout the structure, most notably in the PO$_{4}$ polyhedra, as seen in Fig.~\ref{fig2}(b). I classify the distortion through analysis of the symmetry-adapted phonon modes before and after Cu substitution finding the structural distortion to be caused by $\Gamma_1$ and  $\Gamma_2$ modes of amplitudes 1.19~\AA\ and 1.78~\AA, respectively. Visualization of these symmetry-adapted modes is given in the SI (Fig. S1); this analysis confirms that the structural distortion caused by the Cu substitution on Pb(1) is primarily driven by polyhedral tilts of the PO$_4$ and their corner-shared oxygen neighbors.

Looking closely at the change in coordination of Cu on the Pb(1) site, I see the Cu$^{2+}$ is now six-coordinated with oxygen forming a distorted Jahn-Teller trigonal prism. In an ideal trigonal prism the anions are arranged at the vertices of a regular triangular prism with the transition metal at the center with $D_{3h}$ symmetry. However, breaking in-plane mirror symmetry results in Jahn-Teller distorted trigonal crystal field with $C_{3v}$ symmetry, such as is found in the Janus compounds \cite{Er_et_al:2018}. Here I find the mirror is already broken by the next-nearest neighbor P ions -- the Cu-O distance that has nearby P is $2.06$ \r{A}, whereas the Cu-O  without surround P ions is $2.35$ \r{A}. Such out-of-plane asymmetry in the Cu environment results in a local dipole in $z$ which can also influence the resulting electronic structure \cite{Er_et_al:2018}. In addition to the mirror symmetry breaking, the O triangles are also rotated at a small angle of $\sim24^{\circ}$ with respect to each other, referred to as a Bailar twist in molecules\cite{Bailar_et_al:1964}, giving the final distorted trigonal prism. Interestingly, the same structural distortion has been observed upon substitution of U on the Ca(1) site in fluorapatite Ca$_{10}$(PO$_4$)$_6$F$_2$ as measured from X-ray absorption spectroscopy~\cite{Rakovan_et_al:2002}. There the authors found that the U substitution on the Ca(1) sites causes an usual 6-coordinated structure -- a trigonal prism with six equal bondlengths of 2.06\AA\, with an undetermined Bailar twist angle, however.

\subsection{Electronic Structure}

\begin{figure}
\includegraphics[width=0.49\textwidth]{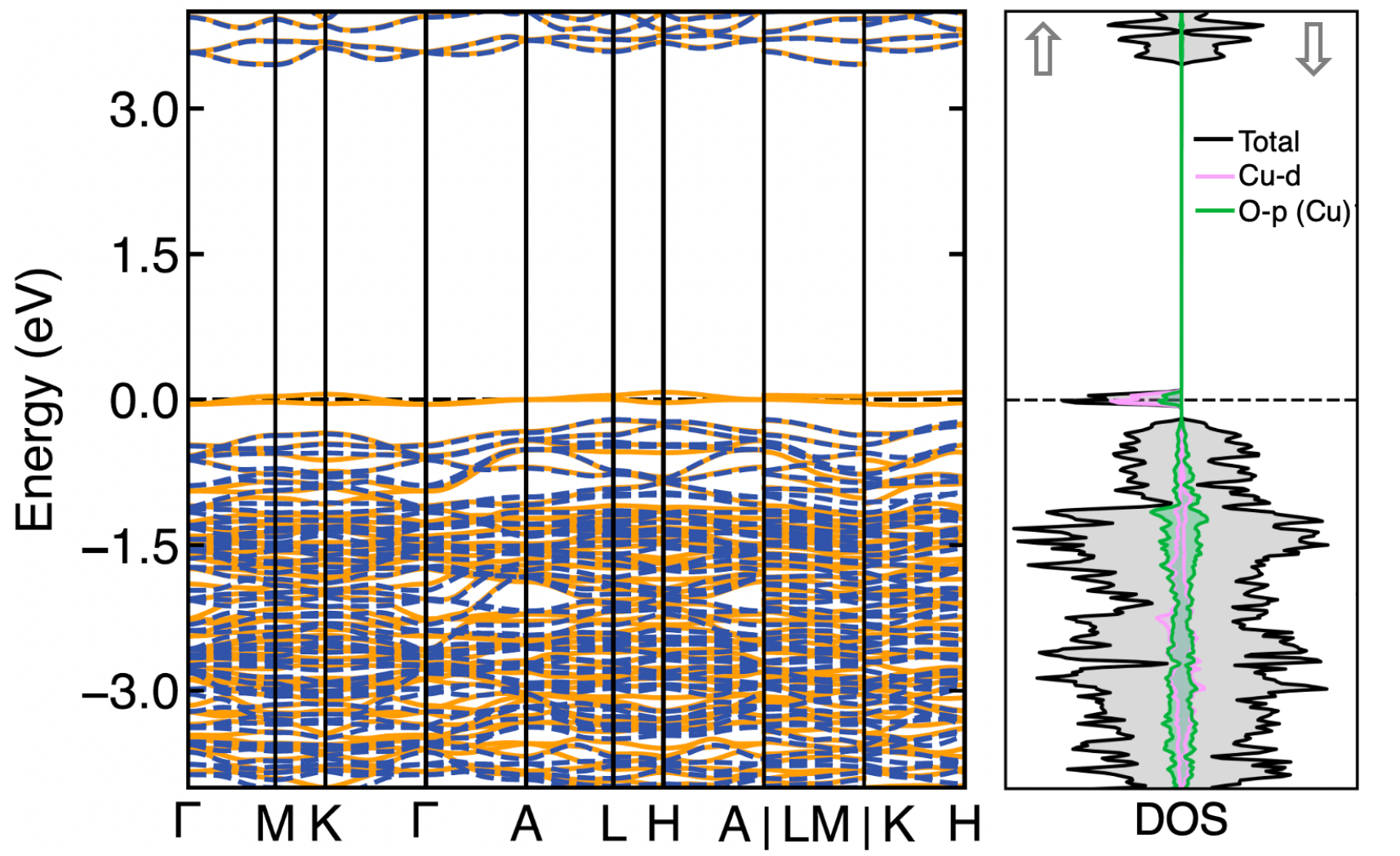}
\caption{Calculated spin-polarized electronic band structure (left) and corresponding density of states (right). The spin-up bands are depicted in solid orange, and the spin-down bands are dashed blue. The total density of states is shaded grey with projections shown of the Cu-$d$ orbitals (pink) and its neighboring O-$p$ orbitals (green). In both plots the Fermi level is set to 0 eV and is marked by the dashed line. } 
\label{fig3}
\end{figure}

\begin{figure}
\includegraphics[width=0.45\textwidth]{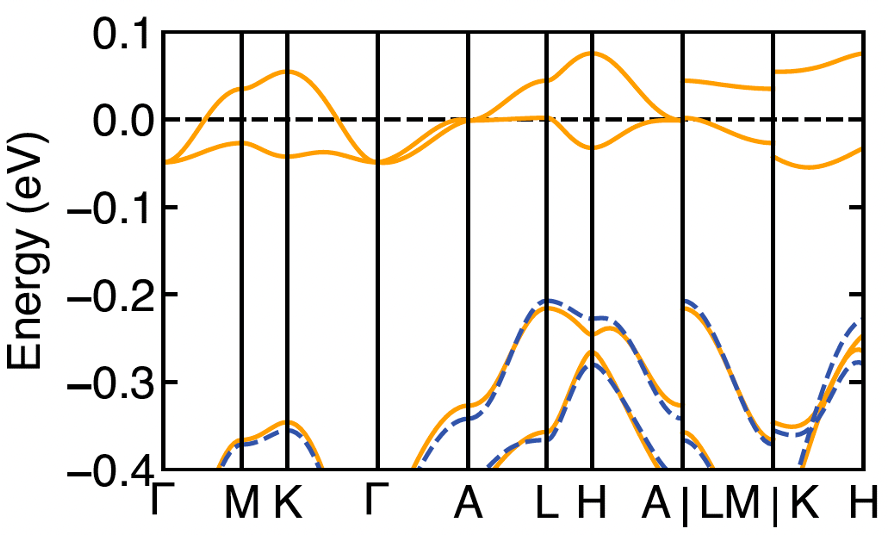}
\caption{Calculated spin-polarized electronic band structure in smaller energy range around the Fermi level showing the isolated two-band Cu-$d$ manifold. The Fermi level is set to 0 eV and is marked by the dashed line. } 
\label{zoom}
\end{figure}

I present the calculated spin-polarized electronic structure in Fig.~\ref{fig3}. Remarkably, I find an isolated set of flat bands crossing the Fermi level, with a maximum bandwidth of $\sim$130 meV (see Fig.\ref{zoom}) that is separated from the rest of the valence manifold by 160 meV. Such a narrow bandwidth is particularly indicative of strongly correlated bands. These especially flat bands are consistent with the Cu-O coordination where I find Cu-O bondlengths of 2.35 \AA\ and 2.06 \AA\ in the distorted trigonal prism. For comparison, the Cu-O bondlengths in the cuprate superconductors are typically $\leq$ 2\AA\ (in-plane) and  $\leq$ 2.3\AA\ (apical)~\cite{Li_et_al:2019}, giving further evidence of the unusual coordination and resulting band localization in this isolated Cu-$d$ manifold.

The crystal-field splitting for the trigonal prism comprises a single $d_{z^{2}}$, doubly degenerate $d_{xy}$ and $d_{x^{2}-y^{2}}$, and doubly degenerate $d_{yz}$ and $d_{xz}$. In fact, the relative positioning of the $d_{z^{2}}$ and the $d_{yz}/d_{xz}$ is set by the anisotropy induced by the mirror-symmetry breaking -- large values of asymmetry can cause the $d_{z^{2}}$ to be destabilized. In our case for Cu$^{2+}$ with a $d^{9}$ configuration I expect half filling of the doubly degenerate $d_{yz}/d_{xz}$ bands -- this is corroborated with our calculations in Fig.~\ref{fig3} where I find two bands of $d_{yz}/d_{xz}$ character at the Fermi level that are are half filled. My results suggest that low-energy physics of this system can be described by a two-band $d_{yz}/d_{xz}$ model, similar to that originally suggested for Fe-pnictide superconductors \cite{Han_et_al:2008, Raghu_et_al:2008}. However, unlike other correlated-$d$ band superconductors, in this system the Cu-$d$ bands are particularly flat -- there is minimal band broadening from neighboring oxygen ions. If previous assumptions about band flatness driving superconductivity are correct, then this result would suggest a much more robust (higher temperature) superconducting phase exists in this system, even compared to well-established high-T$_C$ systems.

While the calculations presented here are performed for GGA+U with a U = 4 eV applied to the Cu-$d$, I find the same qualitative features of the band structures for a wide range of U values, as presented in the SI. This suggests that the presence of such an isolated manifold of flat Cu-$d$ bands is a feature of the structural reconstruction and new crystal field environment provided by the apatite network. I further confirm this by calculating the band structure of the Cu-substituted compound without performing a structural relaxation, that is, taking the structure of the ideal Pb$_{10}$(PO$_{4}$)$_{6}$(OH)$_{2}$ and replacing a Pb(1) with Cu without allowing the structure to relax. In this case, I find slight spin-polarization in the band structure, however all Cu-$d$ states are now in the bulk valence manifold and do not form an isolated manifold (see SI Fig.~S3). Finally, the top of the valence band is made up of the Pb(2)-s states, confirming their contribution to the stereochemical activity of the lone pairs (see SI). 

I next investigate the exchange interactions between Cu ions in different unit cells by constructing doubled unit cells in the in-plane and out-of-plane directions. I find that there is a preference of 2 meV/Cu for ferromagnetic coupling in the out-of-plane direction (with a Cu-Cu separation of $c=7.307$ \r{A}), whereas a preference of 7 $\mu$eV/Cu for antiferromagnetic coupling in the in-plane direction (with a Cu-Cu separation of $a=9.738$\r{A}). However, while this result is indicative of the potential exchange interactions in the system, it relies on the unrealistic assumption that the Cu ions will sit on the same substitution position in each unit cell. A full study of potential exchange interactions for various Cu locations is out of the scope of this work, and will crucially be informed by the Cu locations determined in experiment. 

So far all calculations have been for Cu on the Pb(1) site, which is the site occupancy reported in Ref.~\cite{RoomT}. I also calculated the structural and electronic properties of Cu in the Pb(2) site with the resulting structure and bands given in the SI. In this location, the Cu interrupts the hexagonal network that is characteristic of the apatite structure, which causes a structural rearrangement to the lower $P1$ symmetry, which would have a profound effect on structural measures such as x-ray diffraction. The Cu substituted in this position is now tetrahedrally coordinated by oxygen, and has a significantly different electronic structure with no correlated $d$ bands crossing the Fermi level, as detailed in the SI. In fact, I find that Cu on this Pb(2) is 1.08 eV more energetically favorable than Cu on the Pb(1) site, suggesting possible difficulties in robustly obtaining Cu substituted on the Pb(1) site.

\section{Discussion}
These theoretical results suggest that the apatite structure provides a unique framework for stabilizing highly localized Cu-$d^9$ states that form a strongly correlated flat band at the Fermi level. The central role of stereochemically active 6s$^{2}$ lone pairs of Pb(2) manifests in the formation of a chiral charge density wave and the propagation of structural distortions with connected polyhedra. When Cu is substituted on a Pb(1) site, the result is a cascade of structural alterations, including reduced lattice parameters, changes in coordination, and modified polyhedral tilts, leading to a local Jahn-Teller distorted trigonal prism around Cu. This results in an unusually flat set of isolated $d_{yz}/d_{xz}$ bands with half-filling. 

I briefly note that achieving such a crystal field environment should also be possible in intercalated twisted
heterogeneous bilayers where selection of different heterobilayers can provide the mirror symmetry breaking,
while moir\'{e} twist can provide an arbitrary rotation of the
upper and lower triangles. In fact such a platform would
be ideal for probing the physics found here given its broad
range of tunability and the state-of-the-art characteriza-
tion probes for their interrogation~\cite{VanWinkle_et_al:2023}.

I now discuss the potential implications of these features of Cu-substituted apatite for possible high-T$_C$ superconductivity, and in particular the role of the flat bands and the presence of competing magnetic interactions and phonons. Isolated, flat bands have long been a target for achieving high-T$_C$ as predicted from BCS theory\cite{Heikkila_et_al:2011,Kopnin_et_al:2011,Hofmann_et_al:2020}. The critical temperature T$_C$ given by by T$_C \propto\exp(-1/|U|\rho_0(E_F))$, where $|U|$ is the magnitude of the effective attractive interaction and $\rho_0(E_F)$ is the density of states at the Fermi surface. In a flat band where the density of states diverges, T$_C$ is proportional to $|U|$. This suggests that, at low interaction strengths, the critical temperature in flat bands could be significantly enhanced, as is currently being explored in various moir\'{e} systems\cite{Torma_et_al:2022}. In our case, the correlated bands have a bandwidth of $\sim$ 130 meV which is less than their separation from the rest of the valence manifold (160 meV) at the level of density functional theory. While strong correlations will need to be taken into account adequately to accurately describe these scenarios, these are promising hints for these proposals that predict that the T$_C$ is proportional to the interaction strength.

In contrast to conventional superconductors, where phonon-mediated interactions govern Cooper pair formation, there is wide consensus that other bosonic excitations are responsible for the attractive pair formation in high-$T_C$ superconductors ranging from paramagnons to spin- and charge-density waves~\cite{Bertel_et_al:2016}. In this system, I have identified several potential sources of fluctuations that could contribute to pairing. Firstly, I identified a charge density wave (CDW) driven by chiral lone pair ordering on the Pb(2) sites -- the presence of this CDW is strongly connected to the structural rearrangement that occurs when Cu is incorporated into the Pb(1) lattice sites. In addition to this, I identified two zone-center phonon modes that trigger the global structural deformation that occurs as a result of the Cu substitution, suggesting potentially strong electron-phonon coupling for these modes. Finally, I calculated the relative exchange interactions between Cu in neighboring unit cells. Interestingly for the out-of-plane coupling, that is, along the Cu-Pb-Cu one-dimensional chains, I find ferromagnetic coupling is favored by 2 meV/Cu over antiferromagnet coupling, even though the Cu are over 7 \AA\ apart, suggesting that spin fluctuations could also play a key role. 

Finally, the calculations presented here suggest that Cu substitution on the appropriate (Pb(1)) site displays many key characteristics for high-T$_C$ superconductivity, namely a particularly flat isolated $d$-manifold, and the potential presence of fluctuating magnetism, charge and phonons. However, substitution on the other Pb(2) does not appear to have such sought-after properties, despite being the lower-energy substitution site. This result hints to the synthesis challenge in obtaining Cu substituted on the appropriate site for obtaining a bulk superconducting sample. Nevertheless, I expect the identification of this new material class to spur on further investigations of doped apatite minerals given  these tantalizing theoretical signatures and experimental reports 
 of possible high-T$_C$ superconductivity.

\paragraph*{Acknowledgements-.} 
I am grateful to David Prendergast, Adam Schwartzberg, Shuhada' Sadaqat and John Vinson for insightful discussions and encouragement. I also thank D. Kwabena Bediako, Donny Evans, Katherine Inzani, Adam Schwartzberg and John Vinson for feedback on this draft, and Jeff Neaton and Richard Martin for further comments. This work was funded by the U.S. Department of Energy, Office of Science, Office of Basic Energy Sciences, Materials Sciences and Engineering Division under Contract No. DE-AC02-05-CH11231 within the Theory of Materials program. Computational resources were provided by the National Energy Research Scientific Computing Center and the Molecular Foundry, DOE Office of Science User Facilities supported by the Office of Science, U.S. Department of Energy under Contract No. DEAC02-05CH11231. The work performed at the Molecular Foundry was supported by the Office of Science, Office of Basic Energy Sciences, of the U.S. Department of Energy under the same contract.

\clearpage

\section{Supplemental Material}
\subsection{Calculation Details}

I used the Vienna Ab initio Simulation Package (VASP) \cite{Kresse1993,Kresse1994,Kresse1996,Kresse1996a} for all density functional theory (DFT) calculations with projector augmented wave (PAW) pseudopotentials \cite{Blochl1994,KresseA} including Pb 5d$^{10}$6s$^{2}$6p$^{2}$, Cu 3d$^{10}$4s$^{1}$, P 3s$^{2}$3p$^{5}$, O 2s$^{2}$2p$^{4}$, and H 1s$^{1}$ as valence electrons. I used a plane wave cut-off energy of 600 eV and a $6\times6\times8$ Gamma-centered k-point grid for structural optimizations and a $10\times10\times12$ Gamma-centered k-point grid for the density of states and electron localization functional calculations. All calculations are done using the generalized gradient approximation (GGA) based exchange-correlation functional PBEsol \cite{Perdew2008}, with a Hubbard-U of 4 eV applied to the Cu-d states as implemented with the Dudarev approach unless otherwise stated~\cite{Dudarev1998}. This gives lattice parameters within 1\% of experiment~\cite{RoomT} as summarized in Table~\ref{table-structure}. The electronic convergence criterion is set to $10^{-6}$ eV and the force convergence criterion is set to 0.01 eV / \AA{}. We included spin-orbit coupling self-consistently in the electronic structure calculations where specified. Electronic structure plots were generated using the sumo software package~\cite{sumo}.


\subsection{Structural Distortion}

We used the AMPLIMODES software from the Bilbao Crystallographic Server\cite{amplimodes} to analyze the symmetry-adapted phonon modes in going from the stoichiometric Pb$_{10}$(PO$_{4}$)$_{6}$(OH)$_{2}$ with space group $P6_{3}$ to the Cu-substituted compound with Cu on the Pb(1) site, resulting in space group $P3$. A summary of the resulting distortion is given in Table~\ref{amplimodes} with a visualization of the two modes, $\Gamma_1$ and $\Gamma_2$ depicted in Fig.~\ref{figs1}.

\begin{table*}[]
\begin{tabular}{|ccccc|}
\hline
\multicolumn{1}{|l|}{\textbf{Irrep}} & \multicolumn{1}{l|}{\textbf{K-vector}} & \multicolumn{1}{l|}{\textbf{Direction}} & \multicolumn{1}{l|}{\textbf{Subgroup}} & \multicolumn{1}{l|}{\textbf{Amplitude (\AA)}} \\ \hline
\textbf{$\Gamma_1$}     & (0,0,0)    & (a)     & $P6_3$     & 1.192        \\ \hline
\textbf{$\Gamma_2$}      & (0,0,0)      & (a)        & $P3$        & 1.778    \\ \hline
\end{tabular}
\caption{Symmetry-adapted distortion information in going from stoichiometric Pb$_{10}$(PO$_{4}$)$_{6}$(OH)$_{2}$ to Cub$_{9}$(PO$_{4}$)$_{6}$(OH)$_{2}$ where Cu is on the Pb(1) site.}
\label{amplimodes}

\end{table*}

\begin{figure*}[!htb]
\includegraphics[width=0.9\textwidth]{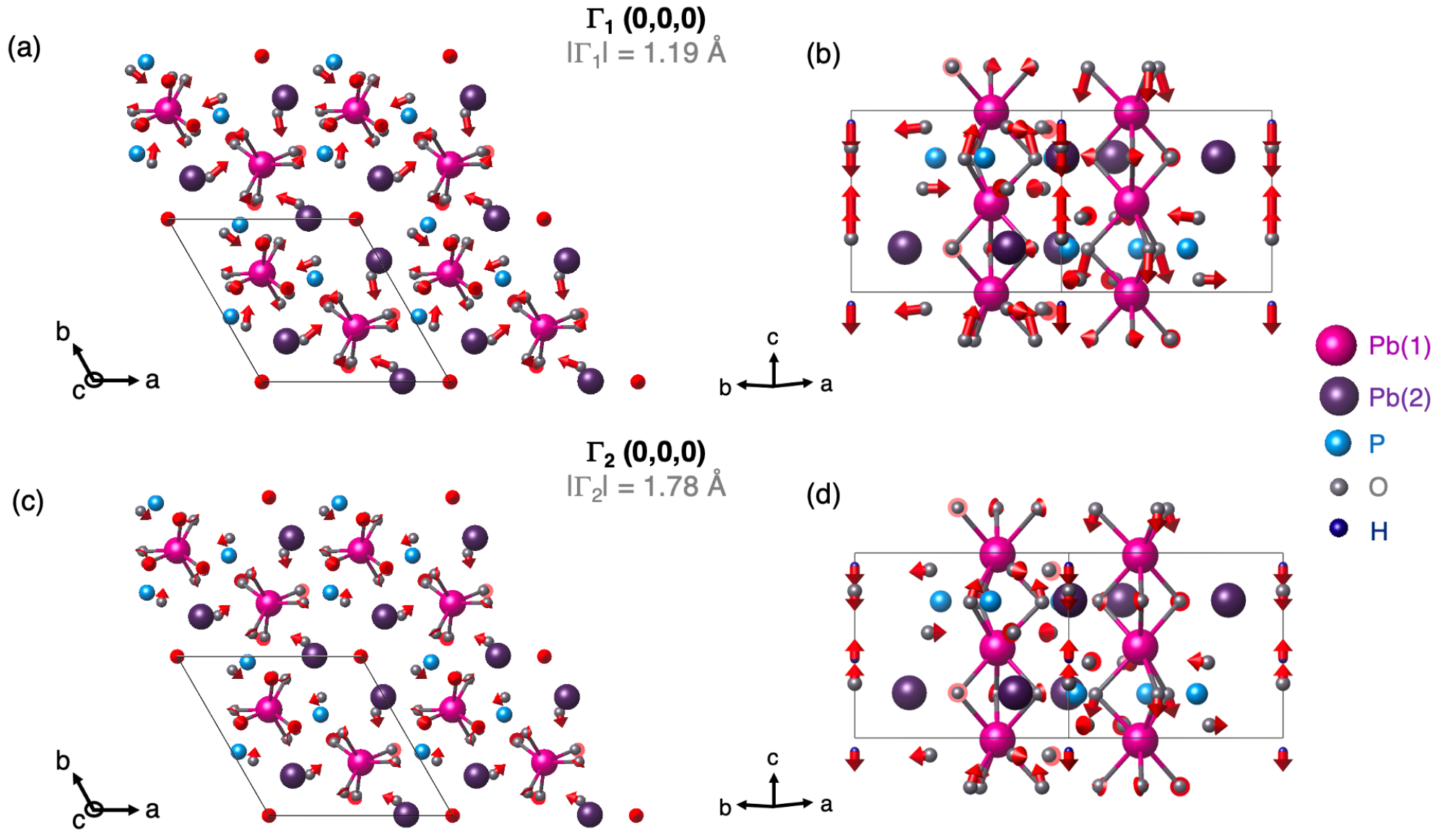}
\caption{Summary of displacements of symmetry-adapted modes upon Cu substitution on the Pb(1) site. (a) and (b) show the $\Gamma_1$ mode which retains the $P6_3$ space group. (c) and (d) show $\Gamma_2$ mode which reduces the $P6_3$ space group to $P3$. The arrows show the ionic motion under each phonon mode with their length being proportional to the amplitude of the displacement.}
\label{figs1}
\end{figure*}

\begin{figure*}
\includegraphics[width=0.9\textwidth]{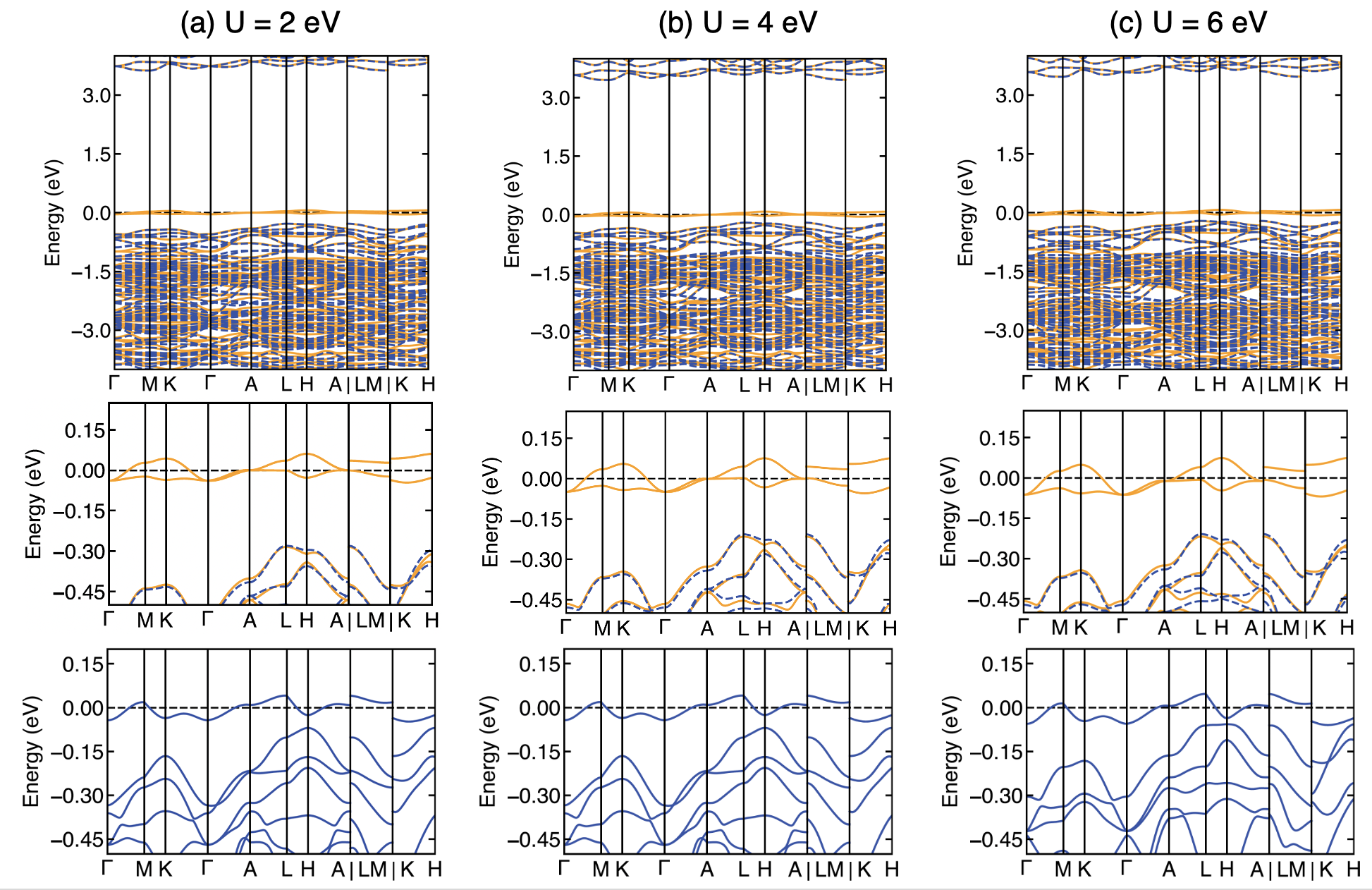}
\caption{(Top panel) Calculated spin-polarized band structure for different values of U on Cu-d states for CuPb$_{9}$(PO$_{4}$)$_{6}$(OH)$_{2}$ with (a) U = 2 eV, (b) U = 4 eV, and (c) U = 6 eV. In all plots the majority spin channel is shown as a solid orange line, and the minority as a dashed blue line. (Middle panel) Zoomed-in version of the top plot of spin-polarized bands with a narrower energy range.  (Bottom panel) Calculated band structures without spin-polarization. The Fermi level is set to 0 eV in all plots and is marked by the dashed line. }
\end{figure*}

\begin{figure*}[!htb]
\includegraphics[width=0.9\textwidth]{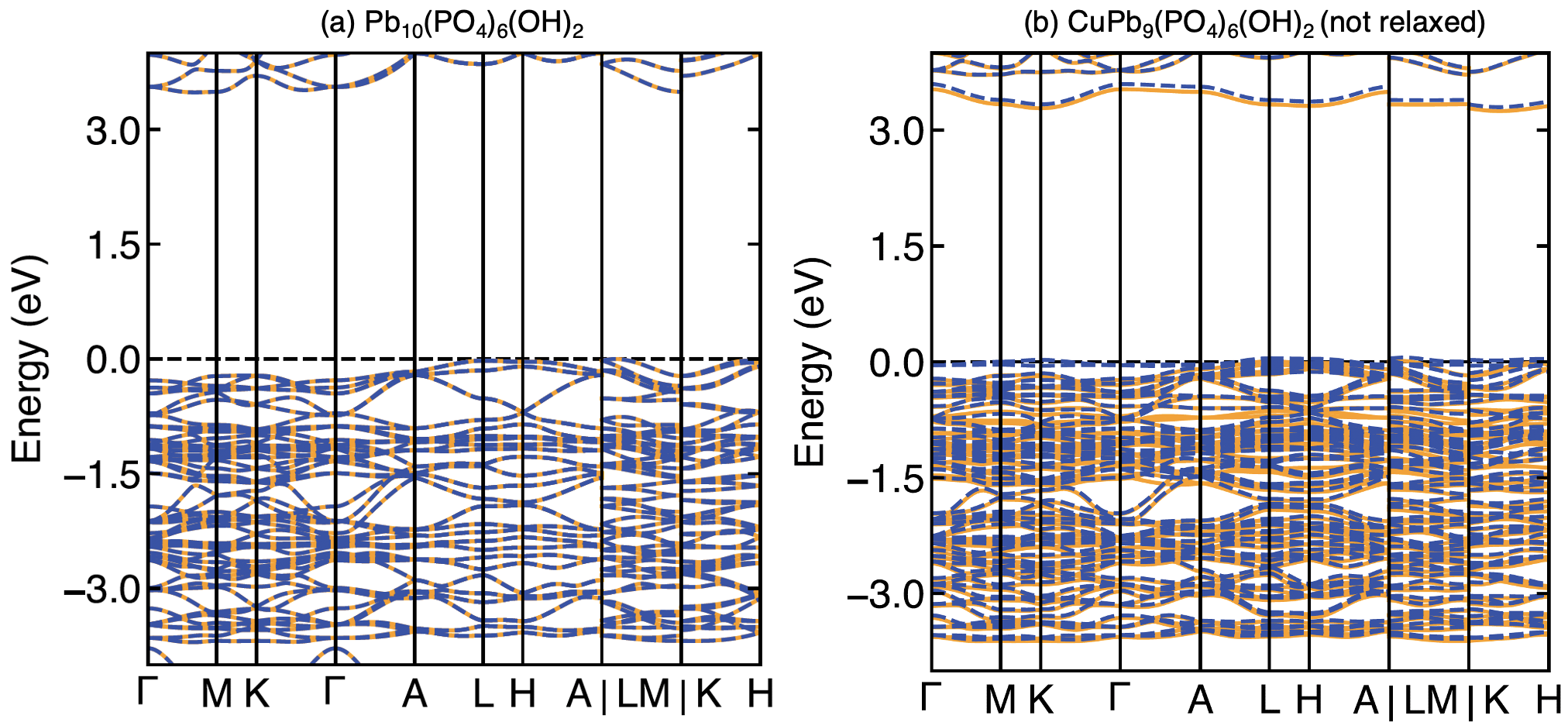}
\caption{Calculated spin-polarized band structures 
 for (a) stoichiometric Pb$_{10}$(PO$_{4}$)$_{6}$(OH)$_{2}$ and (b) CuPb$_{9}$(PO$_{4}$)$_{6}$(OH)$_{2}$ without structural optimization, that is, with the structure of (a) with one Cu replaced for Pb(1). In all plots the majority spin channel is shown as a solid orange line, and the minority as a dashed blue line. The Fermi level is set to 0 eV in all plots and is marked by the dashed line. }
\end{figure*}

\begin{figure*}[!htb]
\includegraphics[width=0.6\textwidth]{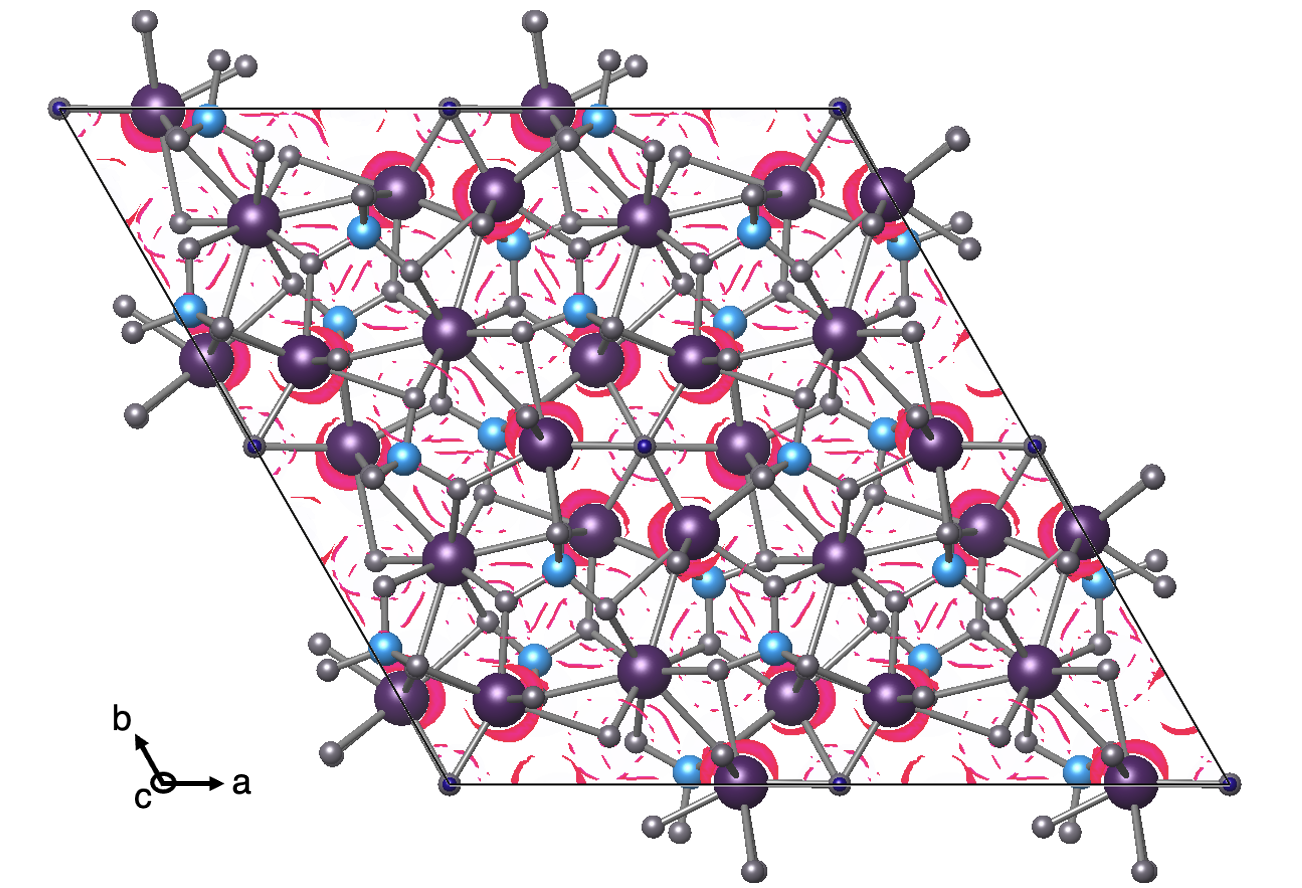}
\caption{Calculated electronic localization function for a 2$\times$2$\times$1 supercell of Pb$_{10}$(PO$_{4}$)$_{6}$OH$_{2}$ showing chiral charge density wave induced by Pb(2) lone pairs.}
\end{figure*}

\begin{figure*}[!htb]
\includegraphics[width=0.9\textwidth]{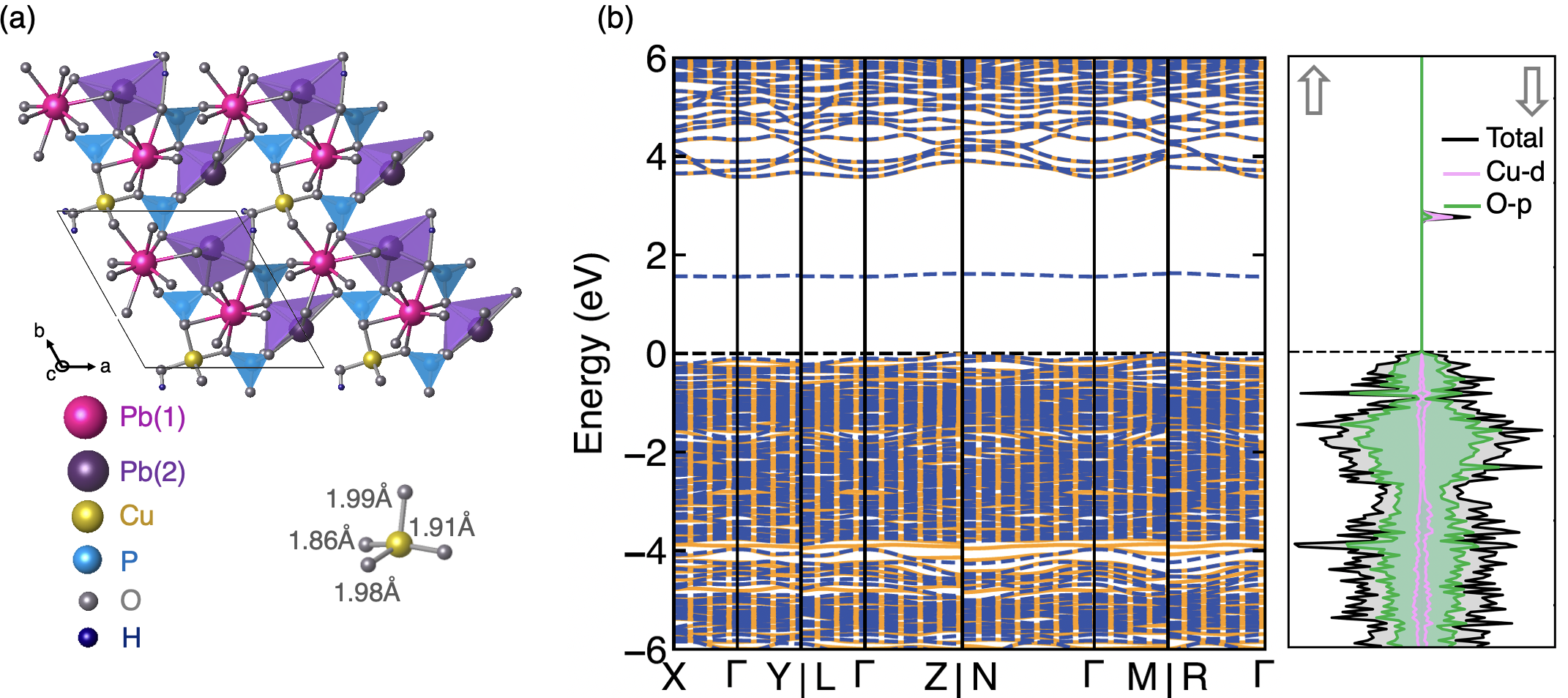}
\caption{Results for CuPb$_{9}$(PO$_{4}$)$_{6}$(OH)$_{2}$ with Cu on the Pb(2) site. (a) The resulting fully relaxed structure for Cu on the Pb(2) site. The inset shows the new local environment of the Cu ion and resulting Cu-O bond lengths. (b) The corresponding calculated spin-polarized electronic structure showing the bands (left) and spin-polarized density of states (right). The total and orbital-projected density of states are shown for the Cu-d and O-p states. The Fermi level is set to 0 eV in all plots and is marked by the dashed line. .}
\label{antisite}
\end{figure*}

\begin{figure*}[!htb]
\includegraphics[width=0.9\textwidth]{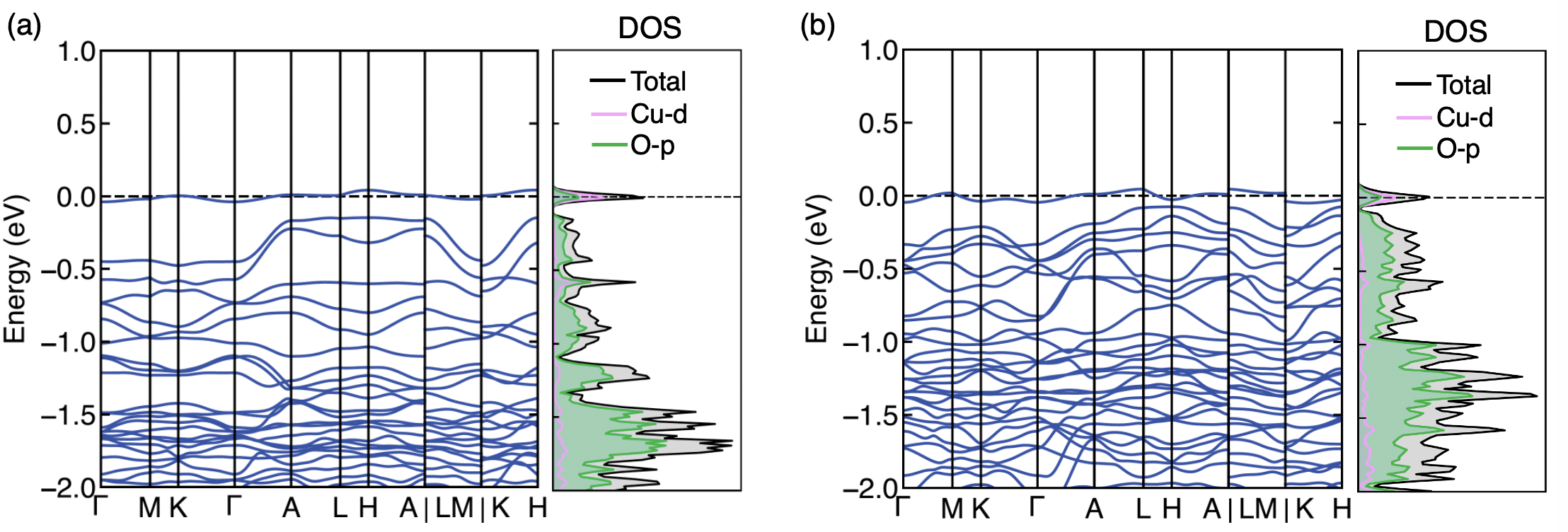}
\caption{alculated non-spin-polarized electronic structures 
 for (a) CuPb$_{9}$(PO$_{4}$)$_{6}$O and (b) CuPb$_{9}$(PO$_{4}$)$_{6}$(OH)$_{2}$. For both calculations the Cu is placed on the Pb(1) site. The total and orbital-projected density of states are shown for replaced for Pb(1). In all plots the majority spin channel is shown as a solid orange line, and the minority as a dashed blue line. The Fermi level is set to 0 eV in all plots and is marked by the dashed line. Note that these were both calculated using the fully relaxed structures that included spin-polarization.}
\label{figs1}
\end{figure*}

\clearpage

\section{Structure Files}
The resulting relaxed structures in POSCAR format are given below. The specific geometry is given in the headers.

\begin{verbatim}
Cu on Pb(1) for OH columns
1.00000
   9.73753   0.00000  -0.00000
  -4.86876   8.43294   0.00000
   0.00000   0.00000   7.30653
Cu Pb P O H 
1 9 6 26 2 
Direct(44) [A1B9C6D26E2] 
   0.33333   0.66666   0.99594  Cu    
   0.66666   0.33333   0.97556  Pb    
   0.66666   0.33333   0.47828  Pb    
   0.33333   0.66666   0.47831  Pb    
   0.24304   0.99176   0.23619  Pb    
   0.74033   0.00357   0.74019  Pb    
   0.00823   0.25127   0.23619  Pb    
   0.99642   0.73675   0.74019  Pb    
   0.74872   0.75695   0.23619  Pb    
   0.26324   0.25966   0.74019  Pb    
   0.39800   0.37662   0.22932  P     
   0.59606   0.63623   0.76062  P     
   0.62337   0.02138   0.22932  P     
   0.36376   0.95982   0.76062  P     
   0.97861   0.60199   0.22932  P     
   0.04017   0.40393   0.76062  P     
   0.33463   0.49455   0.22138  O     
   0.63904   0.50466   0.73061  O     
   0.50544   0.84007   0.22138  O     
   0.49533   0.13438   0.73061  O     
   0.15992   0.66536   0.22138  O     
   0.86561   0.36095   0.73061  O     
   0.58275   0.46627   0.20829  O     
   0.42113   0.55711   0.83037  O     
   0.53372   0.11647   0.20829  O     
   0.44288   0.86401   0.83037  O     
   0.88352   0.41724   0.20829  O     
   0.13598   0.57886   0.83037  O     
   0.32305   0.24965   0.07561  O     
   0.70965   0.76259   0.89995  O     
   0.75034   0.07339   0.07561  O     
   0.23740   0.94705   0.89995  O     
   0.92660   0.67694   0.07561  O     
   0.05294   0.29034   0.89995  O     
   0.61424   0.72712   0.57832  O     
   0.36008   0.28424   0.41289  O     
   0.27287   0.88712   0.57832  O     
   0.71575   0.07584   0.41289  O     
   0.11287   0.38575   0.57832  O     
   0.92415   0.63991   0.41289  O     
   0.00000   0.00000   0.70585  O     
   0.00000   0.00000   0.30956  O     
   0.00000   0.00000   0.44500  H     
   0.00000   0.00000   0.84006  H   
\end{verbatim}
\pagebreak
\begin{verbatim}
Cu on Pb(2) for OH columns
1.00000
9.72342   0.04200   0.00197
-4.82301   8.48804   0.00345
0.00025   0.00382   7.33474
Cu   Pb   P    O    H 
1     9     6    26     2
Direct
0.29671   0.05116   0.33235
0.32381   0.67469   0.98111
0.65139   0.32730   0.98009
0.66758   0.35114   0.48809
0.32169   0.66632   0.47791
0.75782   0.00049   0.75985
0.99392   0.24366   0.24551
0.99739   0.77380   0.76869
0.74898   0.76332   0.23058
0.22801   0.23596   0.76545
0.40018   0.38468   0.22144
0.59137   0.62723   0.73080
0.61312   0.02576   0.23816
0.37685   0.97309   0.72186
0.96697   0.59735   0.22558
0.01790   0.40361   0.72964
0.31214   0.48018   0.22872
0.66266   0.51575   0.73906
0.51223   0.84427   0.23528
0.47793   0.15426   0.72594
0.14692   0.66732   0.22652
0.83604   0.33147   0.73496
0.58332   0.49499   0.22798
0.40826   0.53051   0.72984
0.50719   0.10481   0.25528
0.48254   0.89402   0.72791
0.88433   0.41328   0.23340
0.10201   0.58649   0.73148
0.36287   0.28748   0.04392
0.64492   0.73807   0.90133
0.70876   0.08669   0.05709
0.26165   0.90861   0.88557
0.90639   0.64180   0.05051
0.07638   0.34936   0.89853
0.65158   0.73564   0.56080
0.35772   0.27368   0.38830
0.27627   0.91556   0.54365
0.73438   0.08266   0.39631
0.07010   0.34970   0.55607
0.90705   0.65323   0.39062
0.00689   0.00336   0.92206
0.08251   0.98342   0.30799
0.03614   0.87556   0.35825
0.04711   0.99828   0.04551
\end{verbatim}
\newpage

\begin{verbatim}
Cu on Pb(1) for O columns   
    1.00000     
    9.62586   -0.00000   -0.00000
   -4.81293    8.33624    0.00000
    0.00000    0.00000    7.21588
   Cu   Pb   P    O 
     1     9     6    25
Direct
  0.66667  0.33333  0.98511
  0.00692  0.77593  0.26511
  0.99756  0.23907  0.75708
  0.22407  0.23099  0.26511
  0.76093  0.75849  0.75708
  0.76901  0.99308  0.26511
  0.24151  0.00244  0.75708
  0.33333  0.66667  0.01001
  0.33333  0.66667  0.50100
  0.66667  0.33333  0.49255
  0.62789  0.59460  0.23001
  0.38566  0.39352  0.76637
  0.40540  0.03329  0.23001
  0.60648  0.99214  0.76637
  0.96671  0.37211  0.23001
  0.00786  0.61434  0.76637
  0.49081  0.63441  0.25509
  0.48303  0.30416  0.77686
  0.36559  0.85640  0.25509
  0.69584  0.17887  0.77686
  0.14360  0.50919  0.25509
  0.82113  0.51697  0.77686
  0.75643  0.71787  0.09385
  0.25828  0.33902  0.92184
  0.28213  0.03856  0.09385
  0.66098  0.91926  0.92184
  0.96144  0.24357  0.09385
  0.08074  0.74172  0.92184
  0.29192  0.36023  0.58153
  0.71402  0.61660  0.42013
  0.63977  0.93170  0.58153
  0.38340  0.09742  0.42013
  0.06830  0.70808  0.58153
  0.90258  0.28598  0.42013
  0.56029  0.42167  0.15861
  0.49845  0.57912  0.78156
  0.57833  0.13861  0.15861
  0.42088  0.91933  0.78156
  0.86139  0.43971  0.15861
  0.08067  0.50155  0.78156
  0.00000  0.00000  0.35299

\end{verbatim}
\newpage
\begin{verbatim}
Cu on Pb(2) for O columns   
1.00000
9.78803 0.12357 0.00526
-4.78808 8.26801 -0.00858
0.00359 -0.00485 7.37499
Cu Pb P O
1 9 6 25
Direct
0.95321 0.73690 0.35934
0.00223 0.25142 0.75098
0.21445 0.19820 0.27381
0.74490 0.74531 0.76387
0.77953 0.98644 0.28067
0.25110 0.00434 0.75792
0.68083 0.34384 0.00125
0.32306 0.66524 0.00484
0.33825 0.67242 0.48957
0.65744 0.32831 0.49365
0.61175 0.58921 0.23731
0.37015 0.39196 0.74887
0.41117 0.02923 0.24495
0.60851 0.98175 0.74867
0.97935 0.39884 0.25463
0.02783 0.63040 0.73552
0.50362 0.66504 0.24183
0.47249 0.30930 0.75483
0.34797 0.84661 0.24748
0.69473 0.16747 0.74982
0.16028 0.50174 0.25577
0.84707 0.53213 0.73907
0.70683 0.63213 0.05889
0.26126 0.35126 0.91635
0.34673 0.07836 0.07899
0.65481 0.91395 0.91340
0.92330 0.31167 0.06924
0.09741 0.73767 0.90429
0.25995 0.33621 0.58013
0.72714 0.65832 0.40169
0.65107 0.91793 0.57478
0.36022 0.08525 0.42001
0.08931 0.74346 0.56613
0.92479 0.26882 0.40749
0.51861 0.40431 0.25717
0.47997 0.57775 0.73775
0.59365 0.12144 0.24141
0.42729 0.91309 0.75013
0.90084 0.50623 0.28580
0.09909 0.51545 0.72278
0.99782 0.95063 0.34472
\end{verbatim}

\bibliographystyle{apsrev}
\bibliography{library}

\end{document}